\documentclass[12pt]{iopart}
\usepackage{etoolbox}
\csundef{equation*}
\csundef{endequation*}
\usepackage{amsmath} 

\usepackage[dvips]{graphicx}
\usepackage{iopams}

\usepackage{amsfonts}
\begin{document}
\title[The length scale measurements of the Fractional quantum Hall state on cylinder]
{The length scale measurements of the Fractional quantum Hall state on cylinder}
\author{Qi Li, Na Jiang,  Zheng Zhu and Zi-Xiang Hu }
\address{Department of Physics, Chongqing University,  Chongqing, 401331, P.R. China} 
\ead{zxhu@cqu.edu.cn}

\date{\today}
\begin{abstract}
Once the fractional quantum Hall (FQH) state for a finite size system is put on the surface of a cylinder, the distance between the two ends with open boundary 
conditions can be tuned as varying the aspect ratio $\gamma$.  It scales linearly as increasing the system size and therefore has a larger adjustable 
range than that on disk. The previous study of the quasi-hole tunneling amplitude on disk in Ref.~\cite{Zk2011} indicates that the tunneling amplitudes  have a 
scaling behavior as a function of the tunneling distance and the scaling exponents are related to the scaling dimension and the
charge of the transported quasiparticles.   However, the scaling behaviors poorly due to the narrow range of the tunneling distance on disk.
Here we systematically study the quasiparticle tunneling amplitudes of the Laughlin state in the cylinder geometry which shows a much better scaling behavior.
Especially, there are some corssover behaviors at two length scales when the two open edges are close to each other. These lengths are also reflected in the
bipartite entanglement and the electron Green's function as either a singularity or a crossover.
These two critical length scales of the edge-edge distance, $L_x^{c_1}$ and $L_x^{c_2}$, are found to be related to the dimension reduction and back scattering 
point respectively.

\end{abstract}

\maketitle

\section{Introduction}

The strongly correlated electron system reveals a plenty of non-trivial properties
 which beyond the single-particle picture. 
The fractional quantum Hall effect (FQHE)~\cite{Tsui} is a paradigm of strongly correlated
system that occurs in two-dimensional electron gas with a perpendicular magnetic field. The FQH state is
one of the mostly studied object in the condensed matter physics which has topological protected ground state and non-trivial excitation. 
Especially, the FQH states on the second Landau level, such as $\nu = 5/2$ and  $\nu = 12/5$, are expected to have non-Abelian excitations 
and have potential applications in topological quantum computation~\cite{Kitaev, Freedom, RMPNayak}.
Quasiparticle tunneling through narrow constrictions or point contacts that bring counter-propagating
edges close could serve as a powerful tool for probing both the bulk topological
order and the edge properties of fractional quantum Hall liquids. In
particular, interference signatures from double point contact devices may reveal the statistical
properties of the quasiparticles that tunnel through them~\cite{chamon97}, especially the non-Abelian
ones~\cite{stern06,bonderson06}.   In disk geometry, a quasiparticle can tunnel from the center to the edge by a single particle tunneling 
potential $V_{\text{tunnel}} = V_t\delta(\theta)$ which breaks the rotational symmetry~\cite{HZ2009}.  
The ring shape of the Landau basis wave function with angular momentum $m\hbar$ on disk, i.e., $\varphi_m(r)\sim \textbf{r}^m e^{-|\textbf{r}|^2/4}$ has radius $\sqrt{2m} l_B$.
Therefore the tunneling distance $d$ is tuned by inserting $N_{qh}$ flux quantum, or $N_{qh}$ quasiholes at the center, namely $d/l_B =  \sqrt{2(N_{orb} + N_{qh})} - \sqrt{2N_{qh}}$ where $N_{orb}$ is the number of orbitals. 
The shape of the system evolves from disk to annulus and finally to a quasi-1D ring as increasing $N_{qh}$.
In the ring limit (or CFT limit) with $d \rightarrow 0$, or $N_{qh} \rightarrow \infty$, 
we found a universal analytical formula for the tunneling amplitudes of the bulk quasihole~\cite{Zk2011} and the edge excitations~\cite{Zk2012}. On the other hand, the quasihole
tunneling amplitudes were found to have a scaling behavior as a function of the system size $N_e$ and the tunneling distance $d$. Interestingly, the fractional charge and the scaling dimension appears
in the exponents of the scaling function~\cite{Zk2011}.  However, if we look carefully at the data of the tunneling amplitude in disk geometry, the scaling function works not very well  
for small $d$ which was treated as a finite size effect~\cite{HZ2009} due to a limited number of electrons can be handled in the numerical diagonalization.  It is also hard to look into this region on disk since
$N_{qh} \rightarrow \infty$ while $d \rightarrow 0$.

In this paper, we alternatively consider the physics properties of the FQH liquid in the cylinder geometry. 
The cylinder  has advantages that the distance between the two ends is proportional to the system size and can be tuned from zero to infinity smoothly
by varying the aspect ratio $\gamma = L_y/L_x$, where $L_y$ is the circumference on the side with periodic boundary and $L_x$ is the length of finite cylinder with open boundaries.
As a comparison, in the disk geometry,  the tunneling distance has a maximum which is the radius of the system 
$R = \sqrt{2N_{orb}}$ and $d \sim N_{orb}/\sqrt{2N_{qh}}$ for $N_{qh} \gg N_{orb}$.  This is very inconvenient when we want to look at the small $d$ region since numerous
of quasiholes, or flux will be inserted at the center of disk.
On the other side, when $\gamma \rightarrow 0$,  namely in the thin cylinder limit, two adjacent Landau orbitals have practically 
zero overlap. In this case, the Hamiltonian is dominated by the electrostatic repulsion which contains the direct interaction 
 $\langle \varphi_m \varphi_n |V|\varphi_n \varphi_m \rangle$ and the exchange interaction   $\langle \varphi_m \varphi_n |V|\varphi_m \varphi_n \rangle$. 
 The ground state is generally called a charge density wave (CDW) 
 state, or Tao-Thouless (TT) crystal state~\cite{tao1983, thouless84} on torus
 with the electronic occupation pattern $1001001001\cdots$ in order to minimize the electrostatic repulsion energy.  
The wave function of the FQH state~\cite{EDPRB94} can be obtained by diagonalizing the 
 model Hamiltonian with hard-core interaction, or the Hamiltonian only with
 $V_1 \neq 0$  in the language of the Haldane's pseudopotential.
The more interesting case is when $\gamma \rightarrow \infty$ or $L_y \rightarrow \infty$, to keep the total area of the surface $L_xL_y=2\pi l_B^2 N_{orb}$ invariant, 
or keep the total penetrated flux invariant,  $L_x$ then approaches to zero. It means the two counter propagating edges
at the two ends of cylinder are coming close to each other and the system finally evolves into a one-dimensional system. 
In this case, because of the strongly overlap of all the Landau orbitals,
the Gaussian factors of each Landau wavefunction  are the same and can be erased by normalization. In this one-dimensional limit, the FQH
wave function can be described by the Jack polynomials and therefore 
all the results are the same as that we did on disk in the ring limit. 
The Jack polynomial is one of the polynomial solutions for Calogero-Sutherland Hamiltonian~\cite{Feigin02} which can 
 describe the Read-Rezayi $Z_k$-parafermion states with a negative parameter $\alpha$ and a root configuration (or partition). Jack polynomial is a powerful method in studying the FQHE as it can 
 construct not only the model wave function for Read-Rezayi series~\cite{bernevig08, bernevig08a, bernevig09}, but also the low-lying excitations~\cite{kihoonprb14,boprl12}.  
Another advantage for cylinder geometry is the computational convenient comparing with either the disk or sphere geometries  which was discussed in the 
density matrix renormalization calculation~\cite{ZZ2012}. In this paper, we reconsider the quasiparticle tunneling with cylinder geometry especially in the 
region of small tunneling distance.  Here the quasiparticle can tunnel from the one edge to another as sketched in Fig.~\ref{fig1}. 
Thus the tunneling distance equals to the length $L_x$ of the system.  We find a richer structure in this region and two characteristic length scales appears
not only on the quasiparticle tunneling, but also in the wavefunction overlap, bipartite entanglement entropy and electron Green's function. 

 The rest of this paper is organized  as follows. In section II, we consider the  tunneling amplitude with varying the length of the finite cylinder for $e/3$ and $2e/3$ quasiholes in Laughlin state.
 In section III, the bipartite entanglement entropy, both in orbital space and real space are discussed.
 The results of the electron Green's functions are discussed in section IV and summaries and discussions in section V. 
\section{Quasiparticle tunneling for Laughlin state}
\label{sec:tunneling}
For electrons on a cylinder with circumference $L_y$ in $y$ direction in a magnetic field perpendicular to the surface, the single electron wave function in the lowest Landau level is:
\begin{equation}                                                                                                                                  
\psi_{j}(\vec{r}) \equiv |j\rangle= \frac{1}{\sqrt{\pi^{1/2}L_y}}e^{ik_{y}y}e^{-\frac{1}{2}(x+k_{y})^{2}}                                                             
\end{equation}    
in which $k_{y}=\dfrac{2\pi}{L_y}j $, $ j=0,\pm1,\pm2\cdots$  are the transitional momentum along $y$ direction. 
Here  the magnetic length $l_B = \sqrt{\hbar c/eB}$   has been set to be unit.  For a finite size system, the number of basis states or orbits, $N_{orb}$, equals to 
the number of magnetic flux quantum penetrate from the surface.  Each orbit occupies an area $2\pi l_B^2$. Therefore,
the length in $x$ direction for a finite system is fixed with a given aspect ratio $\gamma$, namely $L_x/l_B = \sqrt{N_{orb} 2\pi /\gamma}$.  

\begin{figure}
\centering
\includegraphics[width=8cm]{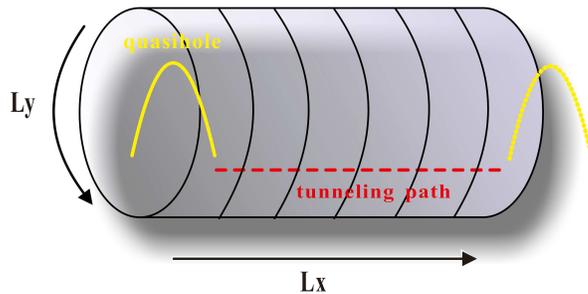}
\caption{\label{fig1}(Color online)The sketch of the quasiparticle tunneling from one side to another side of the cylinder. The tunneling path is given by a potential $V_{\text{tunnel}}=V_t\delta(y)$ which has
length $L_x$.}
\end{figure}
To study the quasiparticle tunneling of the Laughlin state at $\nu=1/3$,  a quasihole with charge $e/3$ or $2e/3$ is put on one edge 
of the cylinder as shown in Fig.\ref{fig1}.   Here the model wavefunction for Laughlin state can be obtained by diagonalizing the model
Hamiltonian with hard-core interaction, or just $V_1 \neq 0$ in the language of the Haldane's pesudopotential. It can also be obtained by using the Jacks with 
the so called root configuration ``$1001001001\cdots$''. The quasihole state for $e/3$ and $2e/3$ is just the translated states with one and two sites along $x$ direction respectively.
Or we can say a quasihole is inserted at the left edge of the finite cylinder which is represented as roots  ``$01001001001\cdots$'' and ``$001001001001\cdots$'' for $e/3$ and $2e/3$ respectively
in the Jack polynomial description.
A simple single particle tunneling potential 
$$V_{\text{tunnel}}=V_t\delta(y)$$ 
is assumed.  It describes a tunneling path along the $x$ direction and therefore  breaks the translational symmetry in $y$ direction.  Then the matrix element is 
$ \langle k \vert V_{\text{tunnel}} \vert m \rangle $,  which is related 
to the tunneling of an electron from the single particle state $\vert m \rangle $   to state $ \vert k \rangle$, is (set $ V_t=1 $)                                                                                                   
\begin{eqnarray}  \label{tunnelingmatrix}                                                                                                                     
v_p(k,m)&=&\langle{k}\vert{V_{\text{tunnel}}}\vert{m}\rangle = e^{-\frac{\pi^2}{L_y^2}(m-k)^2}.                                                                                                                  
\end{eqnarray}  

It is clearly that $v_p(k,m) = e^{-(\frac{2\pi}{L_y}m-\frac{2\pi}{L_y}k)^2/4} = e^{-(d/l_B)^2/4}$ where $d$ is the distance between the two Gaussians.
The many-body tunneling operator can be written as the summation of ones for single particle $\tau = V_t \sum_i \delta(y_i)$. 
Then we can calculate the tunneling amplitude for many-body wave function $\Gamma=\langle\Psi_{qh}\vert \tau \vert \Psi_0\rangle $ 
in which the $\vert \Psi_0\rangle$ and $\vert \Psi_{qh}\rangle$ are the ground state and quasihole state wave function respectively.
In this section, we just consider the tunneling amplitudes for $e/3$ and $2e/3$ quasiholes in Laughlin state at $\nu=1/3$.
The matrix elements consist of contributions from the respective Slater-determinant components $\vert m_1,\ldots,m_N\rangle\in\Psi_{0}$     
and $\vert k_1,\ldots,k_N\rangle\in\Psi_{qh} $. There are nonzero contributions only when the two sets                                
${m_1,\ldots,m_N}$ and ${k_1,\ldots,k_N}$ are identical except for a single pair $m'$ and $k'$ with angular 
momentum difference $k'-m'= N$ for $e/3$ and  $ k'-m'= 2N$ for $2e/3$ where $N$ is the number of electrons.   Therefore, we have 
$v_p^{e/3}=e^{-\frac{\pi^2}{L_y^2}N^2} $ and   $v_p^{2e/3}=e^{-\frac{\pi^2}{L_y^2}(2N)^2}$.  The tunneling amplitude in the second quantization can be written as:
\begin{eqnarray}\nonumber                                                                                                                         
\Gamma &=&\langle \Psi_{qh}\vert \tau \vert \Psi_0\rangle=\sum_i\langle k_1k_2\cdots k_n\vert \delta(y_i)C_{k}^+C_m\vert m_1m_2\cdots m_n\rangle.                                                         
\end{eqnarray}                                                                                                                                    
From the Eq.~\ref{tunnelingmatrix}, it is known that the tunneling amplitude decreases exponentially as increasing the tunneling distance which is proportional to $|m-k|$.  
The distance of the quasiparticle tunneling of mang-body state, or the length of the cylinder $L_x$ is determined by the size of the system and the aspect ratio $\gamma$. 
For a $N$-particle system at fixed filling factor, the $L_x$ can be tuned from $0$ to $\infty$ by changing the aspect ratio $\gamma$. As that on the disk, a numerous of
quasiholes were added at the center which makes the radius changes from $\sqrt{2N_{orb}}$ to $0$.
And with a given $L_y$, the distance between two single particle orbitals on cylinder is a constant which makes the tunneling distance scale proportional to the system 
size, i.e., $d \propto N$ on cylinder comparing with  $d \propto N^{1/2}$ on disk. The linear relation guarantees a smooth change while varying $\gamma$.
\begin{figure}
\centering
\includegraphics[width=10cm]{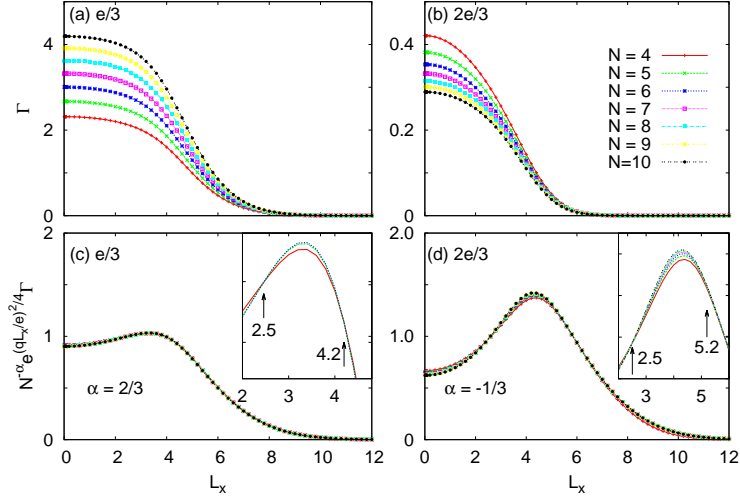}
\caption{\label{tunneling}(Color online) The tunneling amplitude $ \Gamma $  as a function of tunneling distance $L_x$ for $e/3$ (a) and $2e/3$ (b) Laughlin quasihole for system with $4-10$ electrons.
The data from different systems collapse into a single curve with a scaling function in Eq.\ref{scalingansatz}. The exponents are $\alpha^{e/3} = 2/3$ (c) and  $\alpha^{e/3} = -1/3$(d) for two 
types of quasiholes. The insert figure shows the enlarged rescaled data for small $L_x$.  Some crossover points are labeled by arrows.}
\end{figure}

Fig.~\ref{tunneling}(a) and (b) show the tunneling amplitudes for $e/3$ and $2e/3$ Laughlin quasihole as a function of the tunneling distance $L_x$.
When $L_x \rightarrow \infty$,  or $\gamma \rightarrow 0$, the system is in a thin cylinder limit and the ground state is a crystal-like state in which electrons are separated,
then the quasiparticle can not tunnel from one side to another, i.e., $\Gamma \rightarrow 0$.  On the other side, when $L_x \rightarrow 0$, or $\gamma \rightarrow \infty$, all the 
single particle orbitals collapse onto each other which corresponding to the ring limit on disk, or the CFT limit in which case the geometry factor of the many-body wave function can be 
neglected.  Our previous studies~\cite{Zk2011,Zk2012} show that the tunneling amplitude for $e/3$ and $2e/3$ quasihole in the CFT limit for a system with $N$ electrons can 
be exactly represented as:
\begin{eqnarray}\label{CFTvalue}
\Gamma^{e/3}(N) &=& N \frac{\varOmega(1001001...01001)}{\varOmega(01001001...01001)} \nonumber \\
\Gamma^{2e/3}(N)&=&
  2!N\frac{\varOmega\left(\begin{array}{l}
001001001...001 \\
1001001001...001 \\
               \end{array}
   \right)}{\varOmega\left(\begin{array}{l}
01001001001...001 \\
001001001001...001  \\
               \end{array}
   \right)}
\end{eqnarray}
where $\varOmega(1001001...01001) = 1 \times 3 \times 6 \times ...$ and  $\varOmega\left(\begin{array}{c}
\mu \\
\lambda \\
\end{array}
   \right) = \varOmega(\mu)\varOmega(\lambda)$.  For example,  $\Gamma^{e/3}(2) = 2\frac{1 \times 3}{1\times 4} = 1.5$ and $\Gamma^{2e/3}(2) = 2\times 2 \frac{2 \times 3}{1\times 2 \times 4 \times 5} = 0.6$.
   The numbers in the fractions are the position of the $1$s in Eq.~\ref{CFTvalue}.
Formally, the tunneling amplitude for $e/3$ quasihole for Laughlin state has an algebraic expression $\Gamma^{e/3}(N) = \frac{N}{M}B(N,\frac{1}{M})$, in which $M = 1/\nu = 3$ for Laughlin state and 
the beta function $B$ is defined as $ B(x,\beta) = \Gamma(x)\Gamma(\beta)/\Gamma(x + \beta) $.
Fig.~\ref{tunneling}(a) and (b) show that the tunneling amplitudes saturates exactly at these CFT limit values when $L_x \rightarrow 0$.  In the medium region of $L_x$, the tunneling amplitude has a dramatical change from
these CFT values to zero.   The state in this region is close to the Laughlin state,  thus the signal of the decreasing of the quasihole tunneling amplitude can be seemed as a measurement of a phase transition (PT)-like
from the thin cylinder state with zero tunneling amplitude to the CFT limit with a finite tunneling amplitude.  Here we should note that we use the terminology PT-like instead of PT since there is actually no phase transition
in the ground state while varying $L_x$. The topological properties of the ground state in CFT limit are the same as that in the thin cylinder limit ~\cite{Seidel, Bergholtzprb08, Bergholtzprb08a}.
As shown in Fig.~\ref{tunneling}(c) and (d),
the data for different system sizes collapse into each other after the following scaling conjecture is applied
\begin{equation}\label{scalingansatz}
 \Gamma^q(N,L_x) = \Gamma_0 N^{-\alpha^q}e^{(qL_x/2el_B)^2}.
\end{equation}
\begin{figure}
\centering
\includegraphics[width=10cm,height=5cm]{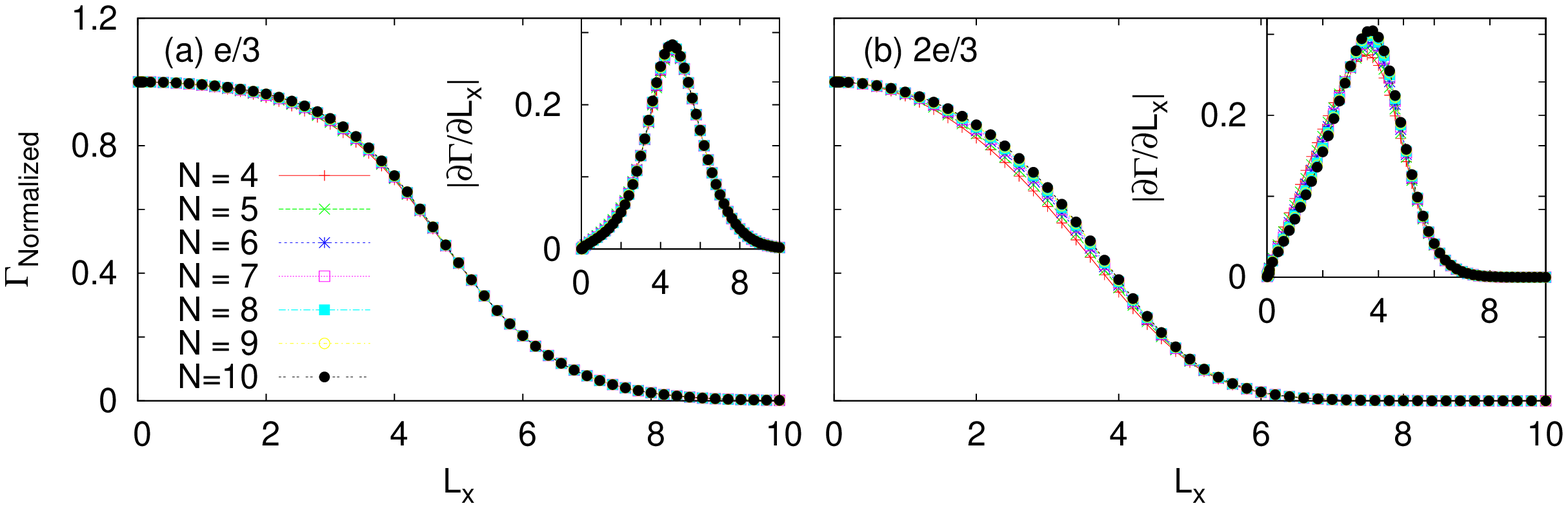}
\caption{\label{fig3}(Color online) (a)The normalized tunneling amplitude as a function of tunneling distance for $e/3$ (a) and $2e/3$ quasihole (b). The insert figures are their first order derivations which
have peaks at  $L_x^{c_2} \simeq 4.2l_B$ both for $e/3$ (a) and $2e/3$(b) by extrapolating to $N \rightarrow \infty$. }
\end{figure}
The exponent $\alpha^q$ is related to the scaling dimension of the quasiparticle as $\alpha^q = 1 - 2 \Delta^q$.  
The $\nu = 1/3$ Laughlin quasihole operator can be written as $\psi_{qh} = e^{i\phi m /\sqrt{3}}$ with a primary charge bosonic field $\phi$ in CFT. Therefore, 
the scaling dimension for $e/3$ and $2e/3$ quasiholes are $\Delta^{e/3} = 1/6$ and  $\Delta^{2e/3} = 2/3$ respectively and then
$\alpha^{e/3} = 2/3$ and $\alpha^{2e/3} = -1/3$. In disk geometry~\cite{HZ2009}, the best scaling parameter for $2e/3$ was $\alpha^{2e/3} = -0.4$ which has a large deviation from 
the theoretical prediction. We think this deviation should come from the insufficient tunneling distance which is maximized at the radius of the disk.  
On the other hand, as we discussed above, the tunnel amplitude data near $d = 0$ is missed due to the huge number of quasiholes needed to be insert
at the center.  On cylinder, as showed in Fig.~\ref{tunneling}(c) and (d), the scaling conjecture of Eq.\ref{scalingansatz} works perfect when $L_x$ is large than a specific value.
The reason we are saying this is that if we enlarge the rescaled data in the small $L_x$ region as shown in the inserted figure, 
 crossover behaviors from different systems occur around $L_x^{c_1} \simeq 2.5l_B$ and  $L_x^{c_2} \simeq 4.2l_B$ as shown by arrows in Fig.~\ref{tunneling}(c). 
 The crossover at $L_x^{c_1} \simeq 2.5l_B$ is also remained the same in Fig.~\ref{tunneling}(d) for $2e/3$ quasihole. Here we should note
that there is no crossover for larger $L_x$ in Fig.~\ref{tunneling}(d). However, we  observe that the scaling behavior is starting to be broken down near $L_x^{c_2'} \simeq 5.2l_B$.  
The first crossover
at $L_x^{c_1} \simeq 2.5l_B$ can be explained a transition from two dimensional system to one dimensional system. The 1D system corresponds to 
Calogero-Sutherland model~\cite{CSM} which actually 
is the origin of the holomorphic part of the FQH wave function, or the Jack polynomials~\cite{bernevig08, bernevig08a, bernevig09}.
The dimension reduction of the FQH state was also considered in the composite ferimion systems~\cite{jainprb90, yuyueprb}.  Or we can say that  the system is in the 
CFT limit while $L_x < L_x^{c_1}$.
The second critical value  $L_x^{c_2} \simeq 4.2l_B$ for $e/3$
and $L_x^{c_2'} \simeq 5.2l_B$ for $2e/3$ is the transition point where the scaling behavior is broken down. This can be explained by
the broken down of tunneling behavior between two independent edges
due to gluing  the two anti-propagating edges together when varying $\gamma$.
Or we can say that the  $L_x^{c_2}$ is the length scale at which two edges start to interact with each other. 
When $L_x < L_x^{c_2}$, there are back scatterings between the two anti-propagating edges. The different
values of $L_x^{c_2}$ between the $e/3$ and $2e/3$ quasiholes should be from the size difference of them.
Another way to extrapolate the critial point  $L_x^{c_2}$ is renormalizing the data by its CFT value from Eq.~\ref{CFTvalue} as shown in Fig.~\ref{fig3}.  Interestingly, the data for different sizes have a scaling-like behavior with 
collapsing onto one curve. The insert plots in Fig.~\ref{fig3} are the first order deviations of the normalized tunneling amplitudes.  Again the first deviations have peaks at  $L_x^{c_2} \simeq 4.2l_B$ both for $e/3$
and $2e/3$ by extrapolating to the thermodynamic limit with $N \rightarrow \infty$.

\begin{figure} 
\centering
\includegraphics[width=8cm]{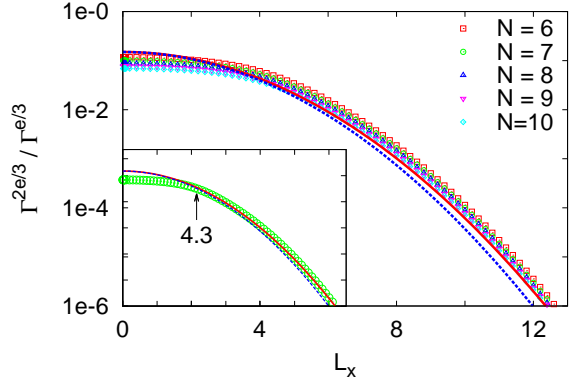}
\caption{\label{fig4}(Color online)The ratio  $\Gamma^{2e/3}/\Gamma^{2e/3} $ as a function of 
$L_x$. The solid line is plotted using Eq.\ref{expectscaling} and he dash line is the fitted line with $\Gamma^{2e/3}/\Gamma^{e/3} \approx e^{-0.078(x/l_B)^2}$.
The insert figure is the enlarged part with smaller $L_x$ for $N=10$ electrons. The numerical data deviates from the asymptotic curve near $L_x^{c_2}\simeq 4.3l_B$. }
\end{figure}
On the other hand, besides the $N$ dependence of the tunneling amplitudes,  the Eq.\ref{scalingansatz} tells us the ratio of the two types of tunneling amplitude is expected to has an asymptotic behavior
which depends on $L_x$:
\begin{equation}\label{expectscaling}
\frac{\Gamma^{2e/3}}{\Gamma^{e/3}} \sim e^{-[(2L_x/3)^2-(L_x/3)^2]/(2l_B)^2} \approx e^{-0.083(L_x/l_B)^2}.
\end{equation}
In Fig.\ref{fig4}, we plot the ratio $\Gamma^{2e/3}/\Gamma^{e/3}$ as a function of $L_x$ for system with $6-10$ electrons. Unlike the data on disk~\cite{HZ2009} in which there was a
sudden change while the first quasihole was inserted at the center, the ratio on the cylinder is smooth as a function of $L_x$. The data in  Fig.\ref{fig4} can be fitted by a solid line with
$\Gamma^{2e/3}/\Gamma^{e/3} \approx e^{-0.078(x/l_B)^2}$ which is consistent with the expected behavior in Eq.\ref{expectscaling}. 
 With comparison to the disk, the ratio of the tunneling amplitude for $e/4$ and $e/2$
for Moore-Read state on disk has an asymptotic $\Gamma^{e/2}/\Gamma^{e/4 } \approx e^{-0.083(d/l_B)^2}$ which has a relative large 
deviation from the expected behavior $\Gamma^{e/2}/\Gamma^{e/4 } \approx e^{-0.047(d/l_B)^2}$.
And for Laughlin state, the numerical and theoretical prediction are  $\Gamma^{2e/3}/\Gamma^{e/3} \approx e^{-0.05(d/l_B)^2}$,
$\Gamma^{2e/3}/\Gamma^{e/3} \approx e^{-0.083(d/l_B)^2}$ respectively.
Moreover, if we just plot the data for 10
electrons as insert plot in Fig.\ref{fig4},  it is shown that the deviation of the asymptotic behavior occurs near $L_x = 4.3l_B$ which is close to $\sim L_x^{c_2}$. 
This deviation also demonstrates that the tunneling has been affected by the edge-edge
interaction at this length scale.

\section{Bipartite Entanglement entropy}
\label{sec:entanglement}
The idea that the quantum entanglement~\cite{Nielsen} in a bipartite system
can describe different phases of matter has emerged over the past years. 
This approach has provided plenty of new insights, while traditional methods based on symmetry breaking and local order parameters in Landau theory are fail. 
More precisely, a bipartition of the quantum system is defined when the Hilbert space factors into two parts $\mathcal{H} = \mathcal{H}_A \otimes \mathcal{H}_B$.  
The bipartite FQH system can be implemented in both the momentum space and the real space of 
the two dimensional electron system. The former is called the orbital cut (OC)~\cite{Haque} and the later real space cut (RC)~\cite{RSES}. 
With a bipartition, a pure quantum state $ \vert \psi\rangle$ can be expressed in the form of Schmidt decomposition
\begin{equation}
\vert \psi \rangle = \sum_ie^{-\xi_i/2}\vert\psi^A_i\rangle \otimes \vert\psi^B_i\rangle,
\end{equation}
where $\vert\psi^A_i\rangle$ and $\vert\psi^B_i\rangle$ are orthonormal sets in $\mathcal{H}_A$ and $\mathcal{H}_B$ respectively and the value of $\xi_i$ in  Schmidt singular
values $e^{-\xi_i/2}$ are the entanglement ``energies'' in entanglement spectrum~\cite{HF2008}. Equivalently, the reduced density matrix  
$\rho_A = \text{tr}_B\vert \psi \rangle {}\langle \psi \vert$ has eigenvalues $\lambda_i = e^{-\xi_i}$. 
The Von Neumann entropy
\begin{equation}
S_A = -\text{tr}[\rho_A\text{ln}\rho_A] = -\sum_i\lambda_i\text{ln}\lambda_i = \sum_i \xi_ie^{-\xi_i} 
\end{equation}
generally scales linearly with the area of the cut between parts A and B and with a universal order $O(1)$ correction, 
namely the topological entanglement entropy~\cite{Pzanardi,Kitaev06,Levin}, i.e., $S = \alpha L - \gamma_t$. 
The topological entanglement entropy $\gamma_t$ of the ground state for a fully gapped Hamiltonian is one robust measure 
of quantum entanglement in a topological phase in two dimension system. In the FQH state, $\gamma_t = \text{log}\mathcal{D}$ where 
$\mathcal{D} \geq 1$ is the total quantum dimension of the system. For Laughlin state at $\nu = 1/3$,  the quantum dimension is $\mathcal{D} = \sqrt{3}$ and
therefore $ \gamma_t \simeq 0.549306$.

\begin{figure}
\centering
\includegraphics[width=10cm]{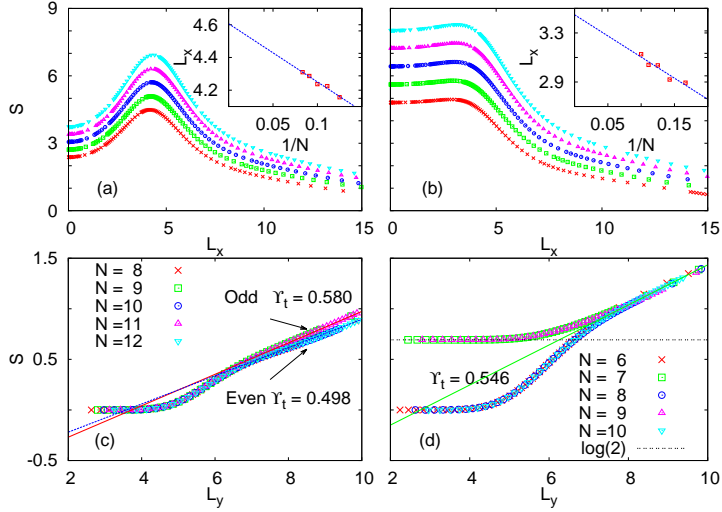}
\caption{\label{fig5} The entanglement entropy as a function of $L_x$ in OC (a) and RC (b) and as a function of $L_y$ in OC (c) and RC (d).  The insert plot in (a) and (b) are 
the scaling of the position of the peak in the entropy. The (c) has even-odd effect in large $L_y$ region and the two fitting straight lines are $0.139L_y - 0.498 \pm 0.001$ (even) and 
$0.155L_y - 0.580 \pm 0.004$ (odd) respectively. The (d) has even-odd effect in small $L_y$ region and the fitting line for large $L_y$ is $0.198L_y - 0.546 \pm 0.006$.  }
\end{figure}

On the cylinder, we intend to divide the system into two equal subsystem with the same
number of orbitals in OC or the same length in RC. However,  the number of orbitals for $N$-electron Laughlin state is $N_{orb} = 3N-2$ which has the same parity as $N$.
Then the bipartition of the orbitals should has an even-odd effect, which can be defined as the orbital difference between $\mathcal{H}_A$ and $\mathcal{H}_B$, namely
$|N_{orb}^A - N_{orb}^B|$ is $0$ for even $N$ and, $1$ for odd $N$.  Intuitively, the effect of orbital difference should be diminished as increasing $L_x$ due to the local
properties of the entanglement entropy, or inversely, it becomes more clear in small $L_x$, or large $L_y$ region. This can be shown in Fig.\ref{fig5}(c).
In the RC case,  it is easy to comprehend that even-odd effect exists, especially in the thin cylinder limit. 
Taking the $N=2$ and $N=3$ thin cylinder crystal-like states as examples, their wavefunction are single Slater-determinant $\Psi_{TT}(2) = |1001\rangle$ and $\Psi_{TT}(3) = |1001001\rangle$ respectively.
Then the position of the RC cut for state with even $N$ has zero electron density which induce a zero entanglement entropy. On the other hand, there is an electron
locates at the position of the RC cut for odd electron TT state. Then the electron density reaches its maximum and the entanglement entropy saturates at a specific value, which
 is the same as that for cutting a single electron wave function into two equal parts which is the classical Von Neumann entropy $S_{TT} = \text{log}(2) \simeq 0.693147$ which is shown
in Fig.\ref{fig5}(d).  When the cylinder is bipartite along the $y$ direction, the $L_y$ is actually the length of the cutting, or the ``area'' between two subsystems.  Then the 
topological entanglement entropy $\gamma_t$ can be extrapolated from Fig.\ref{fig5}(c) and (d).  We found in the case of the OC, the data for systems have the same parity
are sitting on the same curve as a function of $L_y$. For large $L_y$, they can linearly be fitted and the topological entanglement entropy for even and odd parities are 
$\gamma_t \simeq 0.498$ and $\gamma_t \simeq 0.580$ respectively. The exact value $\gamma_t = \log\sqrt{3}$ is in between them.  A more accurate extrapolation can be obtained
in RC entanglement entropy which shown in Fig.\ref{fig5}(d) where the $\gamma_t \simeq 0.546$ which is close to the exact value.  An interesting phenomenon is both the OC and RC
entanglement entropies saturate at zero ($\text{log}(2)$ for odd parity in RC) near $L_y \simeq  L_x^{c_2}$. Therefore, we conclude that $L_x^{c_2}$ is the length scale in $y$
direction that the CDW behavior appears.

In Fig.\ref{fig5}(a) and (b), we plot the bipartite entanglement with OC and RC as a function of $L_x$. The OC entanglement entropy has a peak at $L_x \simeq 4.6 l_B$ in the 
thermodynamic limit as shown in the insert plot. 
In the RC, the peak of the entanglement entropy is not as sharp as that in the OC since
the entropy while $L_x \rightarrow 0$ decreases very slowly. The position of the peak in the thermodynamic limit is $L_x \simeq 3.7  \pm 0.1 l_B$. The error bar origins from the 
strong even-odd effect in this region.  The difference between OC and RC can be explained by the different width of the cuts in realspace. The OC has a wider cut range and is 
more sensitive  to the change of $L_x$.
It is known that the entanglement entropy has a singularity at the critical point of the QPT due to the divergence of the quantum fluctuation. However, since there is no phase 
transition occurs while varying the $L_x$ ~\cite{Seidel, Bergholtzprb08, Bergholtzprb08a}, the increment of the entanglement entropy origins from the correlations between two edges.
Therefore, the two length scales $L_x \simeq 4.6 l_B$ and $L_x \simeq 3.7 l_B$ for OC and RC respectively should be related to the $L_x^{c_2} \simeq 4.2 l_B$.

\section{Electron Green's Function}
\label{sec:Green's function}
\begin{figure}
\centering
\includegraphics[width=8cm]{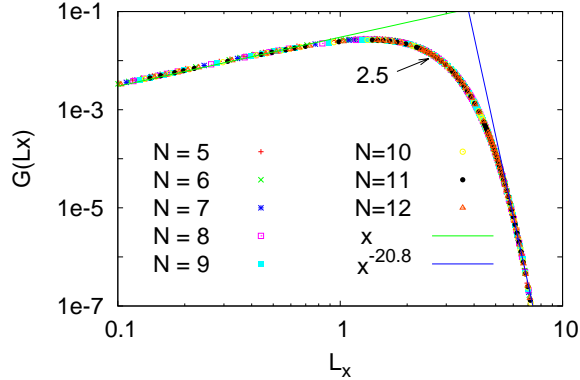}
\caption{\label{fig6}(Color online)The log-log plot of the electron Green's function $\vert G(L_x) \vert$ for the Laughlin state  for 5-12 electrons  as a function of $L_x$. The data in the large and small $L_x$ region
can be fitted by $\sim L_x^{-20.8}$ and $\sim L_x$ respectively. $L_x = 2.5l_B$ is the saddle point between these two phases.}
\end{figure}
Tunneling characteristic at the edge has long been regarded as  an  experimental  method  of  measuring  the  topological order of the FQH liquids. 
For tunneling from a three-dimensional Fermi liquid into the FQH edge,  chiral Luttinger liquid  theory~\cite{wen92}  leads  to  a  non-Ohmic  tunneling
$I-V$ relation $I \propto V^\alpha$ with $\alpha \neq 1$, in sharp contrast to the Ohmic prediction of a Fermi-liquid-dominated edge with $\alpha = 1$. The electron Green's
function is defined as
\begin{equation}
G(\mathbf{r}-\mathbf{r'}) = \frac{\langle\psi\vert\Psi^+_e(\mathbf{r})\Psi_e(\mathbf{r'})\vert\psi\rangle}{\langle\psi\vert\psi\rangle}
\end{equation}
where the $\Psi^+_e(\mathbf{r})$ and $\Psi_e(\mathbf{r'})$ are field operators which create and annihilate an electron at position $\mathbf{r}$ and $\mathbf{r'}$ respectively.  
If we consider the tunneling path $|\mathbf{r} - \mathbf{r'}|$ along the edge of the FQH droplet, the edge Green's function shows a scaling behavior with $\alpha = 1/\nu$ for long distance
tunneling~\cite{xinprl05, hujpcs12}.

In this section, we consider the electron tunneling from the left edge of the cylinder to the right one, namely the electron correlation function between two anti-propagating edges as a function of $L_x$.
The results are shown in Fig.\ref{fig6}.  It shows that the Green's function 
decreases dramatically when $L_x$ is larger than the saddle point which is the one dimensional limit threshold value $L_x^{c_1} = 2.5l_B$.  The data for large
tunneling distance near $L_x \sim 10 l_B$ obeys a power law behavior with an exponent less than $-20$.   In the large $L_x$ limit, obviously, the Green's function is zero in the Tao-Thouless state which is an insulator.
We also checked the $\nu = 1/5$ Laughlin state and found that the electron Green's function has the same power law behavior in this region. 
Thus we think that the exponent in the large $L_x$ region depends on the interaction between electrons.  
The electron Green's function scales as a  $G(L_x) \propto L_x$ 
which has a positive exponent one.  Generally, the electron Green's function at zero temperature decays as $G(r) \sim r^{-1-\alpha}$~\cite{Lutherprb,Theumann} in which $\alpha \geq 0$ and $\alpha/2$ is 
the ``anomalous dimension” of the fermion operators. The case for $\alpha = 0$ corresponds to the normal Fermi liquid and  $\alpha > 0$ is due to the correction of the 
electron-electron interaction which is the behavior of a Luttinger liquid.  On the other hand, when $L_x < L_x^{c_1}$, the system enters into a 
one dimensional phase which is described by the Calogero-Sutherland model.  The reason that the correlation decreases as reducing $L_x$ is due to the repulsion between electrons, or 
strictly speaking, the electron Green's function drop to zero in the 1D limit. 

\section{Summary and discussion}
\label{sec:conclusion}
As a conclusion, we confirm that the quasihole tunneling amplitude in the cylinder geometry obeys the scaling conjecture in Eq.\ref{scalingansatz} and the scaling behavior is much better than that on disk.
  Generally the scaling behavior works well when $L_x > L_x^{c_2}$ where  $L_x^{c_2} \simeq 4.2l_B$
for $e/3$ and $L_x^{c_2} \simeq 5.2l_B$ for $2e/3$ with a difference due to different size of the quasihole. The $L_x^{c_2}$ can be explained as the threshold value of the edge-edge back scattering between two edges.
It appears not only in the quasihole tunneling amplitude calculations, but also in bipartite entanglement entropy. 
Therefore, the $L_x^{c_2}$ is the smallest length scale that guarantees  there are two independent edges at two ends of the cylinder.
It should be the benchmark of the sample size in designing of the experimental setup of the quasiparticle  tunneling and interference~\cite{Willett09,Willett10}.
Moreover, we found another critical value $L_x^{c_1} \simeq 2.5 l_B$,
which is universal for different types of quasiholes. It can be explained as the critical width evolving from a 2D system to 1D system which is described by the Calogero-Sutherland model.  
Bipartite entanglement entropy has a singular behavior near $L_x^{c_2}$ due to a contribution of the edge-edge back scatterings. 
The topological entanglement entropy is extracted from the OC and RC entanglement entropies as a function of $L_y$ in a finite size system.
The $L_x^{c_1}$ plays a role of a saddle point in the single particle Green's function where
the system enters into a one dimensional description. The scaling exponent of the Green's function is one while approaching to the 1D limit.  We notice that the $L_x^{c_1}$
is actually the correlation length of the Laughlin state as mentioned in the iDMRG calculation~\cite{Pollman}.
Here we should admit that we just consider the Laughlin state of the model Hamiltonian with hard-core interaction, or with $V_1$ pesudopotential. 
For a realistic coulomb interaction or the FQH state in higher Landau levels, we believe that the similar behaviors exists which may at most
have small modifications on the value of these lengh scales.

We thank X. Wan and G. T. Liu for helpful discussions. This work was supported by NSFC Project No. 1127403 and Fundamental Research Funds for the Central Universities No. CQDXWL-2014-Z006.  
NJ is also supported by Chongqing Graduate Student Research Innovation Project No. CYB14033.

\end{document}